\documentclass [12pt]{article}
\title{Stationary Statistics of Turbulence as an Attractor}
\author{Edsel A. Ammons\\Department of Physics and Astrophysics\\University of North Dakota\\101 Cornell Street Stop 7129\\Grand Forks, North Dakota 58202}
\date{May 16, 2010}
\begin{document}
\maketitle
\begin{abstract}
A calculational approach in fluid turbulence is presented.  Use is made of the attracting nature of the fluid-dynamic dynamical system.  An approach is offered that effectively propagates the statistics in time.  Loss of sensitivity to an initial probability density functional and generation of stationary statistical effects is speculated.
\end{abstract}
\section{Introduction}
Turbulence of a system is thought to be a condition of the system for which the space and time evolution of the dynamical variables are chaotic, but for which a statistical distribution functional is applicable.  The time evolution of an initially given probability distribution functional for the dynamical variables is given by the deterministic evolution equations that govern the system, called deterministic chaos.

A fluid environment is a dynamical system, having energy input mechanisms at long distance scales, and energy dissipation mechanisms due to viscosity occurring at small distance scales.  Hence, the fluid has an attractor, called the Strange Attractor in the turbulent regime, for its dynamical variables.  These considerations apply to the standard system of interacting, or self-interacting, classical fields which are governed by evolution equations that are at most first order in the time derivatives, and that possess dissipation mechanisms, together with stationary boundary conditions.  It is proposed here that the fluid probability density functional also has an attractor for its time evolution, and an approach to generating this time evolution is presented.  The same mechanism that causes the dynamical variables to have an attractor in phase space, that is, the tendency for the equilibration of energy input rates and energy output rates to set in, also causes an arbitrary initial statistics to evolve toward an attracting statistics, which is stationary in time.

It is the stationary statistics that allow the Kolmogorov scaling ideas to have applicability.  The evolution of the fluid's statistics can be set up as part of a space-time path integral.  Ensemble averages of any dynamical variable can be formulated in terms of this path integral.  Fluid space-time sampling techniques naturally suggest a useful way, using a relatively arbitrary initial statistics functional, to calculate averages.

\section{Description of the Mathematical Approach to Turbulence}
Let us set up the evolution equations for incompressible fluid dynamics.  The extension of the following to compressible fluids will pose no unusual difficulty.

We have,$$\rho \frac{d\vec{v}}{dt}=\vec{f}+\eta \nabla ^{2}\vec{v},$$where $\vec{f} $ is an external force density, such as due to a scalar pressure field, and $\eta $ is the coefficient of viscosity.  $\vec{v} $ is the fluid velocity field.  We also have from the conservation of mass, $$ \nabla \cdot (\rho \vec{v} ) + \frac{\partial \rho }{\partial t}=0. $$  Here, $\rho $ is the mass density of the fluid.  If this mass density is a constant, then the velocity field is divergenceless.  Also, $$\frac{d\vec{v}}{dt}=\frac{\partial \vec{v}}{\partial t} + \vec{v}\cdot \nabla \vec{v}.$$ So we have the fluid dynamic system, \begin{eqnarray} \frac{\partial \vec{v}}{\partial t} & = & -\frac{\nabla P}{\rho } - \vec{v} \cdot \nabla \vec{v} + \nu \nabla ^{2} \vec{v} \label{eq:first}\\ \nabla \cdot \vec{v} & = & 0,\label{eq:second} \end{eqnarray} where $P$ is the pressure, and $\nu \equiv  \frac{\eta }{\rho }.$

We drop the external force density in what follows.  This is not an essential step.  What are needed are a set of interacting fields, together with stationary boundary conditions to allow a deterministic time evolution of the set.  We also associate with the spatial velocity field a probability density functional, $\rho [v,t].$  The fluid statistics time-evolves according to deterministic chaos ~\cite{Ro:Rosen} ~\cite{Th:Thacker},$$\rho [v_{f},t_{f}]=\int d[v_{0} ]K[v_{f},t_{f};v_{0},t_{0}]\rho [v_{0},t_{0}],$$ where the kernel is the delta functional, $$K[v_{f},t_{f};v_{0},t_{0}]=\delta [v_{f}-f[t_{f};v_{0},t_{0}]].$$  That is, the number, $\rho $, associated with the spatial velocity field $v_{0}$ at time $t_{0}$, will be associated with $v_{f}$ at time $t_{f}$, where $v_{f}$ is the velocity field $v_{0}$ deterministically evolves into from time $t_{0}.$  Given a functional of the spatial velocity fields, $A[v]$, its ensemble average, at time $t_{f}$ is, $$<A[v]>=\int d[v_{f}]A[v_{f}]K[v_{f},t_{f};v_{0},t_{0}]\rho [v_{0},t_{0}]d[v_{0}].$$

We want to propagate the fluid's statistics according to deterministic chaos, even though the detailed fluid orbits are chaotic.  Let, \begin{eqnarray*} <A[v]> & = &\int A[v_{f}]\rho [v_{f},t_{f}]d[v_{f}] \\ & = & \int A[v_{f}]K[v_{f},v_{0}]\rho [v_{0},t_{0}]d[v_{f}]d[v_{0}]\\ & = & \int A[f[v_{0}]]\rho [v_{0},t_{0}]d[v_{0}],\end{eqnarray*} where, \begin{eqnarray*} K[v_{f},v_{0}] & = & \delta [v_{f}- f[v_{0}]] \\ & = & \int \delta [v_{f} - f_{1}[v_{1}]]\delta [v_{1} - f_{1}[v_{0}]]d[v_{1}] \\ & = & \int \delta [v_{f} - f_{2}[v_{1}]]\delta [v_{1} - f_{2}[v_{2}]]\delta [v_{2} - f_{2}[v_{0}]]d[v_{1}]d[v_{2}] \\ & = & \int \delta [v_{f} - f_{3}[v_{1}]]\delta [v_{1} - f_{3}[v_{2}]]\delta [v_{2} - f_{3}[v_{3}]]\delta [v_{3} - f_{3}[v_{0}]]d[v_{1}]d[v_{2}]d[v_{3}] \\ & = & \cdots ,\end{eqnarray*} where the velocity fields, $v_{1}$, $v_{2}$, $v_{3}$, ... occur in chronological order, $v_{1}$ being closest to the time $t_{f}.$  Eventually, we have an $f_{M},$ where $M$ is large, such that $v_{M}=f_{M}[v_{0}]$ is infinitesimally different from $v_{0}.$

Hence, $$<A[v]>=\int d[v]A[v_{f}]\delta [v - f_{M}[v]]\rho [v_{0},t_{0}],$$ where the functional integration measure is, $$d[v]=d[v_{f}]d[v_{1}]\cdots d[v_{M}]d[v_{0}],$$ and $$<A[v]>=\int A[f_{M}[\cdots [f_{M}[f_{M}[v_{0}]]]\cdots ]]\rho [v_{0},t_{0}]d[v_{0}].$$

The exact rule, $f_{M}[v],$ requires a solution of the fluid dynamic equations, incorporating the boundary conditions.  The exact rule, $f_{M}[v],$ is difficult to find. Let us use an approximate rule, $F_{M}[v].$  Then, we may say, with motivation to follow, \begin{eqnarray} <A[v]> &= & \int d[v]A[v_{f}]\delta [v-F_{M}[v]]\lambda [\nabla \cdot v]\lambda [v - v_{B}]\rho [v_{0},t_{0}]. \end{eqnarray}  The functional integration is over all space-time velocity fields within the spatial system, between times $t_{0}$ and $t_{f}.$  $\delta [v - F_{M}[v]]$ is a space-time delta functional.  $F_{M}[v]$ generates $v$ from a $v$ an infinitesimal instant earlier.  The $\lambda $ functionals are evaluated at a particular instant.  The needed properties of the $\lambda $ functionals are $\lambda [0]=1,$ and $\lambda [g \neq 0]=0.$  $\lambda [g] $ could be, $$\lim_{\epsilon _{1} \rightarrow 0^{+}} e^{-g^{2}/\epsilon_{1}}.$$  We have, then, \begin{eqnarray} <A[v]> & = & \int d[v_{0}]A[F_{M}[F_{M}[\cdots [F_{M}[v_{0}]]\cdots ]]]\rho [v_{0},t_{0}] \nonumber \\ & & \lambda [F_{M}[\cdots [F_{M}[v_{0}]]\cdots] - v_{B}]\cdots \lambda [v_{0} - v_{B}]  \nonumber \\ & & \lambda [\nabla \cdot F_{M}[ \cdots [F_{M}[v_{0}]]\cdots ]]\cdots \lambda [\nabla \cdot F_{M}[v_{0}]] \lambda [\nabla \cdot v_{0}]. ~\label{eq:average} \end{eqnarray}  The right-hand side of (\ref{eq:average}) equals $<A[v]>$, because $\rho [v_{0},t_{0}]$ will be non-zero only for spatial fields $v_{0}$ that make all the $\lambda$'s equal one.  This means that we only utilize divergenceless spatial velocity fields satisfying, also, the spatial boundary conditions.

The spatial velocity fields have an attractor, determined by the stationary boundary conditions on the fluid \cite{Ru:Ruelle}.  When the boundary conditions allow steady laminar flow to become established, the attractor consists of a single spatial velocity field.  When the Reynolds number becomes large enough, bifurcations set in, or the onset of instability occurs, and the attractor begins to consist of more than one spatial velocity field.  In the turbulent regime, the attractor consists of many spatial velocity fields, and the fluid accesses these fields according to a probability distribution \cite{Fe:Feigenbaum} \cite{Ka:Kadanoff} \cite{Ka:Chaos}.

Given a functional of the spatial velocity fields, $A[v]$, and the fluid dynamic system of equations, (\ref{eq:first}), (\ref{eq:second}), we will say that its ensemble average when the system has reached its attractor is, \begin{eqnarray} <A[v]> & = & \lim_{t_{f}-t_{0}\rightarrow \infty} \int d[v]A[v_{f}]\delta [v - F_{M}[v]] \lambda [\nabla \cdot v] \lambda [v - v_{B}] \nonumber \\ &  & \cdot \rho [v_{0},t_{0}]\label{eq:path}. \end{eqnarray}  The delta functional condition, $\delta [v - F_{M}[v]]$ implements equation (\ref{eq:first}).  $F_{M}[v]$ is an approximate rule for carrying $v$ from, an earlier instant to a later instant.  In the first approximation, $$F_{M}[v]=v - \Delta t (\vec{v} \cdot \nabla v -\nu \nabla ^{2} v),$$ where $v$ is evaluated at an instant $\Delta t$ earlier.  $F_{M}[v]$ can be expressed in all higher powers of $\Delta t$, with only higher order spatial derivatives being required.  This is due to the fact that the hydrodynamic evolution equations are, at most, first-order in the time derivatives.  $\lambda [\nabla \cdot v]$ implements a zero divergence condition on the spatial velocity fields, and $\lambda [v - v_{B}]$ requires the spatial velocity fields to have values $v_{B}$ on the spatial boundaries.

Let us consider the path integral (\ref{eq:path}) to be on a space-time lattice.  We could use, $$\delta (x)= \lim_{\epsilon \rightarrow 0^{+}} \frac{e^{-x^{2}/\epsilon }}{\sqrt{\pi \epsilon }}.$$  We have for the average of $A[v]$ in the steady-state (attractor), letting $\epsilon_{1} (\Delta x)^{2}=\epsilon $, where $\epsilon _{1}$ is in the $\lambda $-functional for the zero divergence condition, and $\Delta x$ being the Cartesian spatial distance step size, \begin{eqnarray} <A[v]> & = & \lim_{\epsilon \rightarrow 0^{+}} \lim_{t_{f} - t_{0} \rightarrow \infty } \int d[v]e^{-H[v]/\epsilon } A[v_{f}]\rho [v_{0}].\label{eq:path2}\end{eqnarray}  Also, $$\rho [v_{0}] \equiv \rho [v_{0},t_{0}],$$ and the functional integration measure, $d[v],$ is, $$d[v]=(\frac{1}{\sqrt{\pi \epsilon}})^{3N(T-1)} \prod_{ijkl} dv_{ijkl}.$$  $H[v]$ is a functional of the lattice space-time velocity field.  Also, neglecting boundary effects, \begin{eqnarray*} H[v] & = & \sum ((v_{l}-v_{l-1}-g[v_{l-1}]\Delta t)^{2} + (v_{x,ijkl} - v_{x,i-1,jkl} + \cdots )^{2})\\ & + & \sum^{'} (v_{ijkl} - v_{B,ijkl})^{2}, \end{eqnarray*} where $$ g[v_{l}]=-\vec{v}_{l} \cdot \nabla v_{l} + \nu \nabla ^{2} v_{l},$$ or, \begin{eqnarray*} g[v_{l}] & = & -v_{x,ijkl}\frac{(v_{ijkl} - v_{i-1,jkl})}{\Delta x}+ \cdots  +\nu \frac{(v_{ijkl}-2v_{i-1,jkl}+v_{i-2,jkl})}{(\Delta x)^{2}}+\cdots. \end{eqnarray*}  $N$ is the number of spatial lattice points in the system, and $T$ is the number of time slices.  We have $\sum $ as a sum over all space-time lattice points in the system, and $\sum^{'}$ as a sum over all space-time lattice points on the spatial boundaries.  Also, $ijk$ are spatial lattice indices, and $l$ is the index in the time direction.

This discretization technique is expected to get better as one increases the lattice fineness and makes use of higher powers of the time step and higher order finite difference approximations of the spatial derivatives \cite{Wa:Warsi}.  A good approximation to the attracting statistics as a starting point will shorten the evolution time required for averages occurring in the steady state.  Gaussian statistics, however, could be a good generic starting point.  The path integral (\ref{eq:path2}) can be evaluated with Monte Carlo techniques utilizing importance sampling.  A calculation of the stationary spatial velocity field that would exist, for the given boundary conditions, if that field were stable, is expected to be a good starting point from which to begin a sampling of the velocity field space-time configuration space.  The average values, for the starting Gaussian statistics, can be taken as the values of the velocities of the calculated stationary spatial velocity field.

We have, $$<A[v]>=J\sum_{i=1}^{n}\frac{A[v_{f,i}]\rho [v_{0,i}]}{n},$$ where, $$J=\int e^{-H[v]/\epsilon }d[v],$$ and the space-time configurations are distributed according to the weighting $e^{-H[v]/\epsilon }.$  $n$ is the number of space-time configurations in the sample.  $A[v_{f,i}]$ is the observable $A[v]$ evaluated at the final time slice of the $i^{th}$ space-time configuration.  The value $\rho [v_{0,i}]$ is attached to the initial time slice of the $i^{th}$ configuration.  We need, $$1=J\sum_{i=1}^{n} \frac{\rho [v_{0,i}]}{n}.$$  So, we must do the integral, $J,$ and constrain $\sum \rho [v_{0,i}]$ to equal $n/J,$ possibly by varying the variances in the Gaussians of $\rho [v,t_{0}].$

\section{Summary}
We have said that the time evolution of the statistics also has an attractor for the following reasons; (1) It is a way to get a solution to the problem of arriving at the steady state turbulence statistics.  One knows that the steady state statistics is stationary with respect to its time-evolver.  Postulating that this statistics is the attractor of its time-evolver means one does not have to have the statistics to get the statistics, thereby offering an approach to the closure problem for the determination of correlations.

(2) The statistical physics approach has been successful in equilibrium thermodynamics, where the derivation of the microcannonical ensemble can be taken as the indication that equilibrium is the attractor for the dynamical system when there is a Galilean frame for which the boundary conditions allow no energy input into the system.  In the attractor, the mean energy input is equilibrated with the mean energy output, because in the attractor dissipative losses have effectively shut off, and the system becomes effectively conservative.  That is, the system becomes describable by a time-independent Hamiltonian. The stationarity of the statistics requires the vanishing of the Poisson bracket of the statistics with this Hamiltonian, resulting in the statistics of equal a priori probability.

(3) The dynamical system, of which a fluid is an example, has an attractor \cite{Ru:Ruelle}.  The dynamics of the statistical approach should mirror the dynamics of the actual dynamical system.

(4) The statistics of the dynamical system, prior to reaching the attractor, has no reason to be unique.  The statistics of the attractor is expected to be unique, in which the geometry of the system, the stationary boundary conditions, and viscosities, all of which determine the Reynolds number, play a crucial role in determining the attractor.

(5)  When the stationary statistics of the fluid is reached, the equilibration of energy input and energy output rates has set in \cite{Fr:Frisch}.

\section{Conclusions}
In the discretized version of the path integral that attempts to arrive at the stationary statistical effects in the generation of the ensemble average of an observable, using an approximate rule for the dynamics, one should arrive at a greater insensitivity to a starting statistics, and a generation of stationary statistical effects, to within any desired level of approximation.  One is trying to use the turbulent transience to get at steady state turbulence effects.  These stationary statistical effects become the backdrop for Kolmogorov's ideas of self-similarity and the resulting scaling exponents.

The potential usefulness of lower dimensional models should not be underestimated \cite{Lo:Lorenz}.  However, it is hoped for, here, that a transition to a detailed use of the Navier-Stokes equations can be brought about.

\section{Acknowledgments}
I wish to acknowledge the Department of Chemistry and Physics of the Arkansas State University, and the Department of Physics and Astrophysics at the University of North Dakota for the environments necessary for this work.  I wish to thank Professor Leo P. Kadanoff, Professor Joseph A. Johnson, and Professor William A. Schwalm for informative and useful discussions.

\end{document}